\documentclass[aip,pop,graphicx,reprint]{revtex4-1}

\usepackage{amsmath,amsthm,amssymb,bm}

\draft % marks overfull lines with a black rule on the right

\begin{document}

%\preprint{}

\title{A Lagrangian perspective on the stability of ideal MHD equilibria with flow} %Title of paper



\author{Yao Zhou}
%\thanks{}
\affiliation{Plasma Physics Laboratory, Princeton University, Princeton, New Jersey 08543, USA}
\author{J. W. Burby}
\affiliation{Plasma Physics Laboratory, Princeton University, Princeton, New Jersey 08543, USA}
\author{Hong Qin}
\affiliation{Plasma Physics Laboratory, Princeton University, Princeton, New Jersey 08543, USA}
\affiliation{Department of Modern Physics, University of Science and Technology of China, Hefei, Anhui 230026, China}


\date{\today}

\begin{abstract}
We take a careful look at two approaches to deriving stability criteria for ideal MHD equilibria. One is based on a tedious analysis of the linearized equations of motion, while the other examines the second variation of the MHD Hamiltonian computed with proper variational constraints. For equilibria without flow, the two approaches are known to be fully consistent. However, for equilibria with flow, the stability criterion obtained from the constrained variation approach was claimed to be stronger than that derived using the linearized equations of motion. We show this claim is incorrect by deriving and comparing both criteria within the same framework. It turns out that the criterion obtained from the constrained variation approach has stricter requirements on the initial perturbations than the other. Such requirements naturally emerge in our new treatment of the constrained variation approach using the Euler-Poincar\'e structure of ideal MHD, which is more direct and simple than the previous derivation from the Poisson perspective.

\end{abstract}

\pacs{}% insert suggested PACS numbers in braces on next line

\maketitle %\maketitle must follow title, authors, abstract and \pacs


\section{Introduction}\label{introduction}
The stability properties of an ideal MHD equilibrium determine whether or not such a configuration can be found in nature or reproduced in the laboratory. Therefore, theoretical tools for assessing the stability of a given equilibrium stand as a cornerstone in the theory of magnetized plasma physics. For \emph{absolute equilibria} (zero flow), where all fluid elements are stationary, a particularly well-developed tool is based on the energy principle\cite{Bernstein,Freidberg,Kulsrud,Laval}, which states that an absolute equilibrium will be linearly stable if and only if the perturbed potential energy $\delta^2 W_0(\bm{\xi},\bm{\xi})$ is non-negative definite as a functional of the the fluid displacement $\bm{\xi}$. For \emph{relative equilibria} (non-zero flow), where fluid elements move along a time-independent Eulerian velocity field, tools for determining stability are not as well-developed. Existing stability criteria only provide sufficient conditions for stability. The oldest condition was found by Frieman and Rotenberg\cite{Frieman} from analyzing the linearized equations of motion. More recently, Hameiri\cite{Hameiri} derived a sufficient condition for stability against so-called dyanmically accessible perturbations, which we will discuss at greater length in what follows. 

Methods for deriving these stability criteria fall into two categories. In the first category\cite{Bernstein,Frieman,Freidberg} are those methods based on careful, direct analysis of the linearized MHD equations of motion. These methods proceed by first proving directly that the linearized force operator is symmetric, and then deriving a stability criterion as a result of this property. The second category consists of methods based on energy conservation \cite{Freidberg,Kulsrud,Laval,Almaguer,Isichenko,Hameiri,Hameiri1998}. Because the non-linear MHD equations possess a conserved energy functional, the linearized equations of motion inherit a conserved quadratic functional. This fact can be exploited to indirectly prove that the linearized force operator is symmetric\cite{Kulsrud}. It also immediately leads to a sufficient condition for linear stability because the conserved quadratic functional provides a norm that bounds linear solutions for all time whenever it is positive definite\cite{Holm1985}. 

When applied to absolute equilibria, either type of method eventually leads to the energy principle, which, again, gives a necessary and sufficient condition for linear stability. However, neither approach gives a necessary and sufficient condition for the linear stability of relative equilibria; the power of each type of method is reduced to only providing sufficient stability criteria. For example, Frieman and Rotenberg's condition\cite{Frieman} asserts that a relative equilibrium will be linearly stable if $\delta^2W(\bm{\xi},\bm{\xi})=-\int\bm{\xi}\cdot\mathbf{F}(\bm{\xi})\,\mathrm{d}^3x$ is positive definite. Here, $\mathbf{F}$ is the force operator appearing in the ideal MHD equations linearized about a flowing equilibrium,
\begin{align}\label{Txi}
\rho\ddot{\bm{\xi}}+2\rho \mathbf{v}\cdot\nabla\dot{\bm{\xi}}=\mathbf{F}(\bm{\xi}),
\end{align} 
where $\rho$ is the mass density and $\mathbf{v}$ is the Eulerian velocity field of the equilibrium. The root of the difficulty in proving \emph{necessity} of this condition, a difficulty which is not present when analyzing absolute equilibria, is the presence of a first time derivative of $\bm{\xi}$ in Eq.\,(\ref{Txi}). 

Another interesting subtlety ushered in by the presence of a non-zero equilibrium flow is that the stability criteria derived from the two approaches no longer coincide. Hameiri's energy-based method\cite{Hameiri} for deriving a condition for linear stability of relative equilibria against dynamically accessible perturbations by leveraging the Poisson structure of ideal MHD\cite{Morrison1980}, which is a refined version of Isichenko's approach\cite{Isichenko}, achieves a different result from the much earlier condition derived by Frieman and Rotenberg. The key idea is that the degeneracy of the field-theoretic Poisson bracket implies that the time derivative of a solution to the non-linear MHD equations cannot point in an arbitrary direction in the (infinite-dimensional) MHD phase space; instead the time-derivative must point along the level sets of the Casimir functionals. The so-called dynamically accessible variations\cite{Morrison,Morrison1989,Hameiri} span these special directions in the phase space. Using this idea, Hameiri argued that a sufficient condition for linearized stability against dynamically accessible variations is the positive definiteness of the functional
\begin{align}\label{D2H}
&\delta^2H(\bm{\xi},\beta,\alpha,\bm{\zeta})=\\
&\int[\rho(\delta\mathbf{v}(\bm{\xi},\beta,\alpha,\bm{\zeta})-\mathbf{v}\cdot\nabla\bm{\xi}+\bm{\xi}\cdot\nabla\mathbf{v})^2-\bm{\xi}\cdot \mathbf{F}(\bm{\xi})]\,\mathrm{d}^3x,\nonumber
\end{align} 
where 
\begin{align}
\delta\mathbf{v}(\bm{\xi},\beta,&\alpha,\bm{\zeta})=\\
&\bm{\xi}\times(\nabla\times\mathbf{v})+{(\nabla\times\bm{\zeta})\times\mathbf{B}}/{\rho}+\alpha\nabla s-\nabla\beta.\nonumber\label{VDA}
\end{align}
Here, $(\bm{\xi},\beta,\alpha,\bm{\zeta})$ are known as Clebsch variables\cite{Marsden1984,Morrison} and they parameterize the dynamically accessible variations. Hameiri noticed that the positive definiteness of $\delta^2W$ is sufficient but not necessary for that of $\delta^2H$. He then concluded that his stability criterion is stronger than Frieman and Rotenberg's.

The proof of the validity of Hameiri's stability criterion comes from the general theory of dynamically accessible variations\cite{Morrison,Morrison1989,Hameiri}, which applies to general Poisson dynamical systems. This theory gives a powerful method for finding stability criteria in various finite and infinite dimensional Hamiltonian systems, and its importance cannot be overstated. However, arguments along these lines, and Hameiri's argument in particular, may be difficult to grasp without a familiarity with infinite dimensional Poisson dynamics. Moreover, the physical origin of the Casimir invariants, which plays a critical role in deriving dynamically accessible stability criteria, may seem mysterious when one starts from a generic Poisson formulation. In particular, an obvious relationship between these invariants and symmetry properties of the underlying physical system is lacking. 

In this paper, we will give a new energy-based derivation of Hameiri's stability criterion for relative ideal MHD equilibria that is more direct and tangible than previous derivations. We will do this by way of an approach based on an Euler-Poincar\'e \cite{Holm,book} counterpart to the method of dynamically accessible variations which we developed for this purpose. Then we will derive Frieman and Rotenberg's criterion within the same framework and compare the two stability criteria. This comparison will yield a subtle surprise which could also be deduced from Morrison's description of dynamically accessible variations on the Poisson side\cite{Morrison,Morrison1989};  Hameiri's criterion requires the initial perturbation to preserve the Noether invariants implied by the particle relabeling symmetry \cite{Cotter}, while Frieman and Rotenberg's does not. This means Frieman and Rotenberg's stability criterion is not strictly weaker than Hameiri's. Thus, the significance of this work is threefold; we provide a simplified means for understanding Hameiri's stability criterion; we outline a method for deriving dynamically accessible stability criteria for general Euler-Poincar\'e fluids; and we clarify that Hameiri's stability criterion applies only to dynamically accessible perturbations while Frieman and Rotenberg's criterion applies to a more general class of perturbations. 

We will begin with two sections describing the Lagrangian counterpart to the method of dynamically accessible variations that we have developed. This general, abstract theory, while illuminating, is not strictly necessary to follow our simplified derivation for both stability criteria, and therefore may be skipped by those readers only interested in the difference between the two criteria. In Sec.\,\ref{EP}, following Ref.\,\onlinecite{Holm}, we will briefly review the notion of an Euler-Poincar\'e fluid. In Sec.\,\ref{constraints}, we will study perturbations to relative equilibria that preserve the Noether invariants implied by the particle relabeling symmetry in a general Euler-Poincar\'e fluid. In particular, we will demonstrate (a) how to derive Clebsch variables that parameterize these perturbations and (b) the existence of a quadratic functional conserved under the evolution of these perturbations. Then in Sec.\,\ref{energy} we will derive both stability criteria based on the Euler-Poincar\'e theory. Finally, we discuss the relationship between our derivation and previous ones, as well as the difference between Hameiri's criterion and Frieman and Rotenberg's in Sec.\,\ref{discussion}.

\section{Euler-Poincar\'e Fluids}\label{EP}
In this section, we briefly review the general notion of an Euler-Poincar\'e fluid in order to prepare for a discussion of linearized stability in this setting in the next section. The material in this section comes directly from the definitive Refs.\,\onlinecite{Holm} and \onlinecite{book}, which develop the theory of Euler-Poincar\'e dynamical systems in great detail. For a detailed explanation of how it applies to the Maxwell-Vlasov system see Ref.\,\onlinecite{Cendra}. Also see Ref.\,\onlinecite{Squire} for an application of the theory to gyrokinetics. As a heuristic rule, conservative fluid systems that exhibit some sort of particle-relabeling symmetry tend to exhibit this type of structure. In order to understand our proof of previous stability criteria later in the paper, knowledge of this formalism is not strictly necessary. However, knowledge of the theory illuminates the structure of our proof. 

Consider a fluid confined within a three-dimensional domain $\mathcal{D}\subset\mathbb{R}^3$. Label points in $\mathcal{D}$ with the variable $X\in\mathcal{D}$ and let the time-dependent mapping $\mathbf{x}_t:\mathcal{D}\rightarrow\mathcal{D}$ give the time-$t$ position of a fluid element $\mathbf{x}_t(X)$ assuming it began at $X$ when $t=0$. If we assume, as we will, that the mapping $\mathbf{x}_t$ is smooth with smooth inverse $\mathbf{x}_t^{-1}$ for all $t$ (note that this may not always be true due to phenomena such as cavitation and shock formation), then we may regard the configuration of the fluid as being specified by an element of the diffeomorphism group $Q\equiv\text{Diff}(\mathcal{D})$, which is the set of all smooth mappings $\mathcal{D}\rightarrow\mathcal{D}$ with smooth inverses. 

Let $V=\Omega_k(\mathcal{D})$ be the space of differential $k$-forms over $\mathcal{D}$ with typical element $u$, i.e. the space of degree-$k$ covariant antisymmetric tensor fields. Let $V^*=\Omega_{3-k}(\mathcal{D})$ be the $(3-k)$-forms with typical element $a$. Finally, let $\mathfrak{X}(\mathcal{D})$ be the space of vector fields on $\mathcal{D}$ tangent to $\mathcal{D}$'s boundary with typical element $\mathbf{v}$. Given a Lagrangian function $L:Q\times\mathfrak{X}(\mathcal{D})\times V^*\rightarrow\mathbb{R}$ satisfying the symmetry property \cite{Note1} $L(\mathbf{x}\circ\mathbf{y},\mathbf{v},\mathbf{y}^*a)=L(\mathbf{x},\mathbf{v},a)$ for all $\mathbf{y}\in \text{Diff}(\mathcal{D})$, the associated \emph{Euler-Poincar\'e fluid with advected parameter} $a_o\in V^*$ is a fluid whose paths in configuration space $t\mapsto \mathbf{x}_t$ are extremals of the action functional
\begin{align}
\mathcal{A}(\mathbf{x}_\cdot)=\int_{t_1}^{t_2}L\left(\mathbf{x}_t,\dot{\mathbf{x}}_t\circ\mathbf{x}_t^{-1},a_o\right)\mathrm{d}t
\end{align}
over the space of curves in $Q$ with fixed endpoints.  Note that $\mathbf{x}_\cdot$ denotes the curve $t\mapsto \mathbf{x}_t$, which is distinct from $\mathbf{x}_t\in Q$. 

This variational principle leads to a second-order differential equation governing the evolution of the fluid configuration $\mathbf{x}_t\in Q$. Because $\mathbf{x}_t$ gives the position of every fluid element in $\mathcal{D}$ at time $t$, this is a Lagrangian, as opposed to Eulerian description of the fluid (unfortunately, we have a Lagrangian Lagrangian description of our fluid). The great utility of the assumed structure of an Euler-Poincar\'e fluid, and in particular the symmetry property possessed by the Lagrangian $L$, is the systematic method it offers for passing to an Eulerian description of the fluid, which we now describe.

Define the projection map $\pi:Q\times\mathfrak{X}(\mathcal{D})\times V^*\rightarrow \mathfrak{X}(\mathcal{D})\times V^*$ using the formula $\pi(\mathbf{x},\mathbf{v},a)=(\mathbf{v},\mathbf{x}^{-1*}a)$. Due to the symmetry property $L(\mathbf{x}\circ\mathbf{y},\mathbf{v},\mathbf{y}^*a)=L(\mathbf{x},\mathbf{v},a)$, there is a unique and well-defined functional $l:\mathfrak{X}(\mathcal{D})\times V^*\rightarrow\mathbb{R}$ satisfying $l\circ\pi=L$, i.e.
\begin{align}
l(\mathbf{v},a)=L(\mathbf{x},\mathbf{v},\mathbf{x}^*a).
\end{align}
Using this \emph{reduced Lagrangian}, we can form a constrained variational principle on $\mathfrak{X}(\mathcal{D})\times V^*$ equivalent to the unconstrained variational principle on $Q\times\mathfrak{X}(\mathcal{D})\times V^*$ defined by the action functional $\mathcal{A}$. As proved in Ref.\,\onlinecite{Holm}, the curve $\mathbf{x}_\cdot$ with $\mathbf{x}_{t_1}=e_1$ and $\mathbf{x}_{t_2}=e_2$ is an extremal of the functional $\mathcal{A}$ if and only if it solves the initial value problem $\dot{\mathbf{x}}_t=\mathbf{v}_t\circ\mathbf{x}_t;~\mathbf{x}_{t_1}=e_1$ for a $\mathbf{v}_t$ that solves the constrained variational problem
\begin{align}\label{cvp}
\delta\int_{t_1}^{t_2}\!l(\mathbf{v}_t,a_t)\,\mathrm{d}t=0.
\end{align} 
The precise definition of this constrained variational problem is as follows. Let $\Gamma_t=(u_t,a_t)\in\mathfrak{X}(\mathcal{D})\times V^*$, where $a_t$ is the unique solution to the initial value problem $\dot{a}_t=-\mathfrak{L}_{\mathbf{v}_t}a_t;~a_{t_1}=e_1^{-1*}a_o$. Let $\Gamma_{t,\epsilon}$ define a two parameter curve in $\mathfrak{X}(\mathcal{D})\times V^*$ with $\Gamma_{t,0}=\Gamma_t$, $\Gamma_{t_1,\epsilon}=(\mathbf{v}_{t_1},a_{t_1})$, $\Gamma_{t_2,\epsilon}=(\mathbf{v}_{t_2},a_{t_2})$, and whose variation has the form
\begin{align}\label{constrained_variations}
\delta\Gamma_t\equiv\frac{\mathrm{d}}{\mathrm{d}\epsilon}\bigg|_0\Gamma_{t,\epsilon}=(\dot{\bm{\xi}}_t+\mathfrak{L}_{\mathbf{v}_t}\bm{\xi}_t,-\mathfrak{L}_{\bm{\xi}_t}a_t),
\end{align}
for an arbitrary time dependent vector field $\bm{\xi}_t\in\mathfrak{X}(\mathcal{D})$ with $\bm{\xi}_{t_1}=\bm{\xi}_{t_2}=0$. Then $\mathbf{v}_t$ is a solution to the constrained variational problem (\ref{cvp}) if
\begin{align}
\frac{\mathrm{d}}{\mathrm{d}\epsilon}\bigg|_0\int_{t_1}^{t_2}\!l(\Gamma_{t,\epsilon})\,\mathrm{d}t=0
\end{align}
for any choice of the time-dependent vector field $\bm{\xi}_t$.

From this new perspective, the constrained variational principle determines the dynamics of the Eulerian velocity field $\mathbf{v}_t$ and the detailed fluid configuration $\mathbf{x}_t$ follows as a sort of afterthought; if we wanted to know how each fluid element moves, all that would be necessary is integration of the time-dependent ODE $\dot{\mathbf{x}}_t=\mathbf{v}_t\circ\mathbf{x}_t$. Therefore, we have arrived at an Eulerian description of our fluid; whereas before the fluid configuration $\mathbf{x}_t$ was regarded as the independent variable, now $\mathbf{v}_t$ is the independent variable. In fact, we can write down the PDE, known as the Euler-Poincar\'e equations, that must be satisfied by $\mathbf{v}_t$ explicitly. We have \cite{Holm,book}
\begin{align}
\frac{\mathrm{d}}{\mathrm{d}t}\frac{\delta l}{\delta\mathbf{v}}=-\mathfrak{L}_{\mathbf{v}_t}\frac{\delta l}{\delta\mathbf{v}}+\frac{\delta l}{\delta a}\diamond a_t,\label{TEP}
\end{align}
which follows directly from the constrained variational principle. Also, recall that the evolution of $a_t$ is assumed to be specified by the equation
\begin{align}
\frac{\mathrm{d}}{\mathrm{d}t}{a}_t=-\mathfrak{L}_{\mathbf{v}_t} a_t.\label{Tadvected}
\end{align}
Here the functional derivative $\delta l/\delta\mathbf{v}\in\mathfrak{X}^*(\mathcal{D})$ is a one-form density, while the diamond operator $\diamond$ obeys the formula
\begin{align}
\left\langle\frac{\delta l}{\delta a}\diamond a_t, \bm{\xi}\right\rangle=-\left\langle\mathfrak{L}_{\bm{\xi}}a_t,\frac{\delta l}{\delta a}\right\rangle.
\end{align}
These notations are briefly explained in the appendix for the readers' convenience. For more detailed discussions see Refs.\,\onlinecite{Holm} and \onlinecite{book}. 

Euler-Poincar\'e fluids are a special case of the more general Euler-Poincar\'e dynamical systems developed by Holm, Marsden, and Ratiu in Ref.\,\onlinecite{Holm}. In the general theory, one starts with a Lie group $G$, a vector space $V$, a (left or right) representation of $G$ on $V$, and a (left or right) $G$-invariant Lagrangian function $L:TG\times V^*\rightarrow\mathbb{R}$. $G$ is regarded as the configuration space for a mechanical system while $V^*$ is regarded as a space of parameters. Hamilton's principle applied to the action functional $\mathcal{A}=\int_{t_1}^{t_2}\!L(g,\dot{g},a_o)\,\mathrm{d}t$ then leads to the usual sort of Euler-Lagrange equations on $TG$, although they are parameterized by the variable $a_o\in V^*$. Finally, by leveraging the $G$-invariance of $L$, an equivalent dynamical description is obtained by reducing Hamilton's variational principle to a constrained variational principle on $\mathfrak{g}\times V^*$, where $\mathfrak{g}$ is the Lie algebra of $G$. The paths in $\mathfrak{g}\times V^*$ satisfying the constrained variational principle obey the Euler-Poincar\'e equations. When $G=\text{Diff}(\mathcal{D})$, $V$ is taken to be a space of differential forms, and the (right) representation of $G$ on $V$ is given by pullback, the Euler-Poincar\'e equations are precisely Eqs.\,(\ref{TEP}) and (\ref{Tadvected}).

When applied to continuum theories, the passage from $TG$ to $\mathfrak{g}$ is precisely the passage from a Lagrangian to an Eulerian description. Thus, Holm, Marsden, and Ratiu's theory offers an attractive means for passing to an Eulerian description of a continuum without losing sight of the variational properties of the Lagrangian description. In particular, the symmetry properties of the Lagrangian imply conservation laws for the Eulerian equations via Noether's theorem or the Kelvin Noether theorem, as discussed in Refs.\,\onlinecite{Cotter} and \onlinecite{Holm}. These conservation laws will play a crucial role in the Lagrangian formulation of the method of dynamically accessible variations that we describe and apply in the coming sections. The distinction between these conservation laws and the advected parameters is also key to understanding the subtle difference between the two stability criteria, while in the previous works from the Poisson perspective \cite{Hameiri} all the constants of motion (Casimirs) were treated as the same.

\section{Clebsch Variables and Energy Conservation}\label{constraints}
Now we will use the general Euler-Poincar\'e equations to study linear perturbations about relative equilibria that preserve the Noether invariants implied by the particle relabeling symmetry. By definition, a relative equilibrium $(\mathbf{v}_o,a_o)$ is a stationary solution to the combined system of PDE's defined by equations (\ref{TEP})  and (\ref{Tadvected}), which we will refer to as the Euler-Poincar\'e system. Moreover, as discussed in Ref.\,\onlinecite{Cotter}, the Noether invariants are given by
\begin{align}\label{noether}
\left<\frac{\delta l}{\delta \mathbf{v}},\bm{\eta}_t\right>=const.
\end{align}
where $\bm{\eta}_t$ is the unique solution to the initial value problem $\dot{\bm{\eta}}_t=-\mathfrak{L}_{\mathbf{v}_t}\bm{\eta}_t;~\bm{\eta}_{t_1}=\bm{\eta}_o$, $\bm{\eta}_o$ is any vector field that generates an infinitesimal symmetry of $a_{t_1}$, i.e. $\mathfrak{L}_{\bm{\eta}_o}a_{t_1}=0$, and $t_1$ is the initial time, which we will assume to be zero henceforth. Therefore, we would like to consider dynamical solutions to the Euler-Poincar\'e system that start out infinitesimally close to $(\mathbf{v}_o,a_o)$ and that all have the same values of the Noether quantities given in Eq.\,(\ref{noether}). As we will see explicitly when applying the general theory we are about to develop to ideal MHD, fixing the values of the Noether quantities is the Lagrangian analogue of fixing the values of the Casimirs when working with Poisson systems. In fact, these Noether invariants, together with the advected parameters which are intrinsic to the Euler-Poincar\'e system, appear to be fully equivalent to the Casimir invariants modulo a Legendre transformation, as will be discussed in a forthcoming article. Thus, we are considering the behavior of what Morrison has dubbed dynamically accessible variations in the Poisson context from the Lagrangian perspective. 

The goal of our analysis is twofold. First of all, we would like to find a parameterization of these dynamically accessible linear perturbations that reflects the fact that any solution to the Euler-Poincar\'e system on $\mathfrak{X}(\mathcal{D})\times V^*$ is the projection of a solution to the variational principle defined by $\mathcal{A}$. The parameterizing variables are known as Clebsch variables. Second, we would like to find the equations of motion for these Clebsch variables and an associated conservation law. 

Let $\Gamma_{t,\epsilon}=(\mathbf{v}_{t,\epsilon},a_{t,\epsilon})$ be an $\epsilon$-dependent family of solutions to the Euler-Poincar\'e system with $\Gamma_{t,0}=\Gamma_o=(\mathbf{v}_o,a_o)$ the stationary solution. If we define the operator $\delta=\frac{\mathrm{d}}{\mathrm{d}\epsilon}\big|_0$, then $\delta\Gamma_t=(\delta\mathbf{v}_t,\delta a_t)\in\mathfrak{X}(\mathcal{D})\times V^*$ is a linearized solution to the Euler-Poincar\'e system. It must satisfy the linearized equations of motion:
\begin{eqnarray}\label{TEP1}
\frac{\mathrm{d}}{\mathrm{d}t}\left(\delta\frac{\delta l}{\delta\mathbf{v}}\right)&=&-\mathfrak{L}_{\delta\mathbf{v}_t}\frac{\delta l}{\delta\mathbf{v}}-\mathfrak{L}_{\mathbf{v}_o}\left(\delta\frac{\delta l}{\delta\mathbf{v}}\right)\\
&&+\left(\delta\frac{\delta l}{\delta a}\right)\diamond a_o+\frac{\delta l}{\delta a}\diamond\delta a_t\nonumber\\
\frac{\mathrm{d}}{\mathrm{d}t}\delta a_t &=&-\mathfrak{L}_{\delta\mathbf{v}_t}a_o-\mathfrak{L}_{\mathbf{v}_o}\delta a_t.\label{Tadvected1}
\end{eqnarray}
In addition, $\delta\Gamma_t=(\delta\mathbf{v}_t,\delta a_t)$ must have the form
\begin{align}\label{special}
\delta\mathbf{v}_t&=\dot{\bm{\xi}}_t+\mathfrak{L}_{\mathbf{v}_o}\bm{\xi}_t\\
\label{special1}\delta a_t&=-\mathfrak{L}_{\bm{\xi}_t}a_o,
\end{align}
for some time-dependent vector field $\bm{\xi}_t$. This follows from the fact that for each $\epsilon$, $\Gamma_{t,\epsilon}$ must be the projection of a solution to the original variational principle defined by $\mathcal{A}$. That is, $\Gamma_{t,\epsilon}=\pi(\mathbf{x}_{\epsilon,t},\dot{\mathbf{x}}_{\epsilon,t}\circ\mathbf{x}_{\epsilon,t}^{-1},a_o)$ (incidentally, this also explains the form of the constrained variations in Eq.\,(\ref{constrained_variations})). Finally, because we assume that the Noether invariants (\ref{noether}) evaluated on $\Gamma_{t,\epsilon}$ are independent of $t$ and $\epsilon$, we have
\begin{align}\label{dynamically_accessible}
\left<\delta\frac{\delta l }{\delta\mathbf{v}}+\mathfrak{L}_{\bm{\xi}_t}\frac{\delta l}{\delta \mathbf{v}},\bm{\eta}\right>=0,
\end{align} 
for all vector fields $\bm{\eta}$ generating infinitesimal symmetries of $a_o$, $\mathfrak{L}_{\bm{\eta}}a_o=0$.

These three conditions on $\delta\Gamma_t$ contain a great deal of information, which we will now unravel. Combining Eq.\,(\ref{TEP1}) with Eq.\,(\ref{special}) and using both the identity $\mathfrak{L}_{\mathbf{v}_o}(u\diamond a)=(\mathfrak{L}_{\mathbf{v}_o}u)\diamond a+u\diamond(\mathfrak{L}_{\mathbf{v}_o}a)$ and the equilibrium condition $\mathfrak{L}_{\mathbf{v}_o}\frac{\delta l}{\delta \mathbf{v}}=\frac{\delta l}{\delta a}\diamond a_o$, we obtain
\begin{align}\label{transformed_linear}
\frac{\mathrm{d}}{\mathrm{d}t}\mu_t+\mathfrak{L}_{\mathbf{v}_o}\mu_t=\omega_t\diamond a_o,
\end{align}
where $\mu_t=\delta\frac{\delta l}{\delta\mathbf{v}}+\mathfrak{L}_{\bm{\xi}_t}\frac{\delta l}{\delta\mathbf{v}}$ and $\omega_t=\delta\frac{\delta l}{\delta a}+\mathfrak{L}_{\bm{\xi}_t}\frac{\delta l}{\delta a}$. This equation may be regarded as a linear, inhomogeneous partial differential equation for the one-form density $\mu_t$. It admits a simple formal solution in terms of the the Lagrangian trajectory $\mathbf{y}_t\in Q$ of the equilibrium flow $\mathbf{v}_o$. Indeed, using the fact that $\dot{\mathbf{y}}_t=\mathbf{v}_o\circ\mathbf{y}_t$, the left-hand side of Eq.\,(\ref{transformed_linear}) can be written $\mathbf{y}_t^{-1*}\frac{\mathrm{d}}{\mathrm{d}t}(\mathbf{y}_t^*\mu_t)$, which implies
\begin{align}\label{seventeen}
\mu_t=\mu_0+\left(\int_{0}^t\!(\mathbf{y}_\tau\circ\mathbf{y}_t^{-1})^*\omega_\tau\,\mathrm{d}\tau\right)\diamond a_o.
\end{align}
Moreover, in light of equation (\ref{dynamically_accessible}), $\mu_0=u_0\diamond a_o$ for some $u_0\in V$. Therefore, by Eq.\,(\ref{seventeen}), for all $t$, there is some $u_t\in V$ and some $\bm{\xi}_t\in\mathfrak{X}(\mathcal{D})$ such that \cite{Note2}
\begin{align}
\label{VEP}\delta\frac{\delta l}{\delta\mathbf{v}}&=-\mathfrak{L}_{\bm{\xi}_t}\frac{\delta l}{\delta\mathbf{v}}+u_t\diamond a_o\\
\label{Vadvected}\delta a_t&=-\mathfrak{L}_{\bm{\xi}_t}a_o,
\end{align}
and
\begin{eqnarray}
\frac{\mathrm{d}}{\mathrm{d}t}{\bm{\xi}_t}&=&-\mathfrak{L}_{\mathbf{v}_o}\bm{\xi}_t+\delta\mathbf{v}_t,\label{VEL}\\
\frac{\mathrm{d}}{\mathrm{d}t}u_t&=&-\mathfrak{L}_{\mathbf{v}_o}u_t+\delta\frac{\delta l}{\delta a}+\mathfrak{L}_{\bm{\xi}_t}\frac{\delta l}{\delta a}.\label{DEP}
\end{eqnarray}

Provided that the Lagrangian is non-degenerate, Eqs.\,(\ref{VEP}) and (\ref{Vadvected}) tell us how to complete the first goal of our analysis; these equations identify the appropriate Clebsch variables \cite{Marsden,Morrison}. The idea behind Clebsch variables in general is to represent quantities subject to constraints in a way that automatically satisfies the constraints. For instance, a magnetic field $\mathbf{B}$ is subject to the constraint $\nabla\cdot\mathbf{B}=0$. If $\mathbf{B}$ is expressed as $\nabla\alpha\times\nabla\beta$, then it clearly automatically satisfies the divergence-free constraint. Conversely, locally every magnetic field can be represented in this way. In this case $\alpha,\beta$ are known as (local) Clebsch variables. In the setting of this paper, Clebsch variables should give a representation of linearized solutions to the Euler-Poincar\'e system that guarantee they satisfy the constraint of preserving the Noether invariants (to first order), i.e. Eq.\,(\ref{dynamically_accessible}). To see that Eqs.\,(\ref{VEP}) and (\ref{Vadvected}) do indeed identify Clebsch variables, first it is necessary to define what it means for the Lagrangian to be non-degenerate. $l$ is non-degenerate when the linear operator $\frac{\delta^2l}{\delta \mathbf{v}^2}(\mathbf{v}_o,a_o):\mathfrak{X}(\mathcal{D})\rightarrow\mathfrak{X}(\mathcal{D})^*$ given by the formula
\begin{align}\label{nondegeneracy}
\frac{\delta^2l}{\delta\mathbf{v}^2}(\mathbf{v}_o,a_o)[\mathbf{w}]=\frac{\mathrm{d}}{\mathrm{d}\epsilon}\bigg|_0\frac{\delta l}{\delta\mathbf{v}}(\mathbf{v}_o+\epsilon\mathbf{w},a_o)
\end{align}
is invertible. When the Lagrangian is non-degenerate, Eqs.\,(\ref{VEP}) and (\ref{Vadvected}) can be used to explicitly express $\delta \mathbf{v}_t$ and $\delta a_t$ in terms of $\bm{\xi}_t$ and $u_t$ using the mapping $\mathfrak{C}:\mathfrak{X}(\mathcal{D})\times V\rightarrow\mathfrak{X}(\mathcal{D})\times V^*$ given by
\begin{align}\label{c}
\mathfrak{C}(\bm{\xi},u)=\bigg(\frac{\delta^2 l}{\delta\mathbf{v}^2}^{-1}\left[u\diamond a_o-\mathfrak{L}_{\bm{\xi}}\frac{\delta l}{\delta \mathbf{v}}+\frac{\delta^2l}{\delta a\delta\mathbf{v}}[\mathfrak{L}_{\bm{\xi}}a
_o]\right]\\
,-\mathfrak{L}_{\bm{\xi}}a_o\bigg),\nonumber
\end{align} 
where $\frac{\delta^2l}{\delta a\delta\mathbf{v}}[a^\prime]=\frac{\mathrm{d}}{\mathrm{d}\epsilon}\big|_{0}\frac{\delta l}{\delta \mathbf{v}}(\mathbf{v}_o,a_o+\epsilon a^\prime)$. Indeed, Eqs.\,(\ref{VEP}) and (\ref{Vadvected}) can be written in terms of $\mathfrak{C}$ as 
\begin{align}
(\delta\mathbf{v}_t,\delta a_t)=\mathfrak{C}(\bm{\xi}_t,u_t).
\end{align}
Here $\bm{\xi}$ and $u$ are the Clebsch variables; any $(\delta\mathbf{v},\delta a)$ expressed in the form $\mathfrak{C}(\bm{\xi},u)$ will automatically satisfy Eq.\,(\ref{dynamically_accessible}). Conversely, by the derivation of Eqs.\,(\ref{VEP}) and (\ref{Vadvected}), any $(\delta\mathbf{v}_t,\delta a_t)$ that is a linearized solution preserving the Noether invariants implied by the particle relabeling symmetry must be of the form $\mathfrak{C}(\bm{\xi}_t,u_t)$.

Now we turn to the first half of the second goal of this section; specifying appropriate dynamical equations for the Clebsch variables $\bm{\xi},u$. Whatever these Clebsch variable evolution equations might be, they must have two properties: (a) if $(\bm{\xi}_t,u_t)$ is a solution to the Clebsch variable evolution equations, then $\mathfrak{C}(\bm{\xi}_t,u_t)$ must be a solution to the linearized Euler-Poincar\'e system (b) every linearized solution of the Euler-Poincar\'e system that preserves the Noether invariants must be given by applying $\mathfrak{C}$ to a solution to the Clebsch variable evolution equations. To find such equations, we only have to examine Eqs.\,(\ref{VEL}) and (\ref{DEP}). If every instance of $\delta\mathbf{v}_t$ and $\delta a_t$ appearing in these equations is expressed in terms of $\bm{\xi}_t$ and $u_t$ using $\mathfrak{C}$, then it becomes clear that these equations define a closed system of time evolution equations for $\bm{\xi}_t$ and $u_t$. Moreover, it is easy to check that solutions to this system of equations map to solutions of the linearized Euler-Poincar\'e system, Eqs.\,(\ref{TEP1}) and (\ref{Tadvected1}), under $\mathfrak{C}$ and satisfy Eqs.\,(\ref{special}) and (\ref{special1}). Thus, these evolution equations satisfy property (a). Conversely, the derivation of Eqs.\,(\ref{VEL}) and (\ref{DEP}) given earlier in this section proves that these evolution equations satisfy property (b). Therefore, Eqs.\,(\ref{VEL}) and (\ref{DEP}) are satisfactory evolution equations for the Clebsch variables.

Finally, we turn to finding a quadratic functional preserved by the dynamics of the dynamically accessible perturbations. Because $l$ has no explicit time-dependence, energy is conserved, which is given by
\begin{align}\label{ep_hamiltonian}
h(\mathbf{v},a)=\left<\frac{\delta l}{\delta\mathbf{v}}(\mathbf{v},a),\mathbf{v}\right>-l(\mathbf{v},a).
\end{align} 
Therefore, when this energy functional is evaluated along the $\Gamma_{t,\epsilon}$ introduced earlier in this section, we have $\frac{\mathrm{d}}{\mathrm{d}t}h\circ\Gamma_{t,\epsilon}=0$. Then, assuming partial derivatives in $\epsilon$ and $t$ commute, we have a primitive form of a linearized conservation law:
\begin{align}\label{conserve}
\frac{\mathrm{d}}{\mathrm{d}t}\frac{\mathrm{d}^2}{\mathrm{d}\epsilon^2}\bigg|_{\epsilon=0}h\circ\Gamma_{t,\epsilon}=0.
\end{align}
Using the fact that $\Gamma_{t,\epsilon}$ is a solution to the Euler-Poincar\'e system with fixed values of the Noether invariants associated with the particle relabeling symmetry, $\frac{\mathrm{d}^2}{\mathrm{d}\epsilon^2}\big|_{0}h\circ\Gamma_{t,\epsilon}$ can be expressed in the form
\begin{align}\label{D2HEP}
&\frac{\mathrm{d}^2}{\mathrm{d}\epsilon^2}\bigg|_{0}h\circ\Gamma_{t,\epsilon}=\\
&\left\langle\delta\left(\mathfrak{L}_{\mathbf{v}_{t,\epsilon}}\frac{\delta l}{\delta\mathbf{v}}-\frac{\delta l}{\delta a}\diamond a_{t,\epsilon}\right),\bm{\xi}_t\right\rangle-\left\langle \delta(\mathfrak{L}_{\mathbf{v}_{t,\epsilon}}{a_{t,\epsilon}}),u_t\right\rangle,\nonumber
\end{align}
where, as before $\delta=\frac{\mathrm{d}}{\mathrm{d}\epsilon}\big|_0$. Using $\mathfrak{C}$, this quantity can be expressed entirely in terms of the Clebsch variables $\bm{\xi}_t$ and $u_t$. Thus, the quadratic functional
\begin{align}\label{D2HEP2}
&\delta^2h(\bm{\xi},u)=\\
&-\bigg<\mathfrak{L}_{\delta \mathbf{v}}\frac{\delta l}{\delta\mathbf{v}}+\mathfrak{L}_{\mathbf{v}_o}\left(\frac{\delta^2 l}{\delta\mathbf{v}^2}[\delta\mathbf{v}]\right)+\mathfrak{L}_{\mathbf{v}_o}\left(\frac{\delta^2 l}{\delta a\delta\mathbf{v}}[\delta a]\right)\nonumber\\
&~~~~~~~~-\left(\frac{\delta^2 l}{\delta\mathbf{v}\delta a}[\delta\mathbf{v}]+\frac{\delta^2 l}{\delta a^2}[\delta a]\right)\diamond a_o-\frac{\delta l}{\delta a}\diamond\delta a,\bm{\xi}\bigg>\nonumber\\
&-\left<\mathfrak{L}_{\delta\mathbf{v}}a_o+\mathfrak{L}_{\mathbf{v}_o}\delta a,u\right>,\nonumber
\end{align}
where $(\delta\mathbf{v},\delta a)=\mathfrak{C}(\bm{\xi},u)$, is conserved by the Clebsch variable evolution equations. Therefore, whenever this functional is positive definite, it defines a norm (in fact, an inner product!) on the space of Clebsch variables, $\mathfrak{X}(\mathcal{D})\times V$, that bounds solutions to the Clebsch variable evolution equations for all time. This means positive definiteness of $\delta^2h$ is sufficient to guarantee stability of the Clebsch variable evolution equations \cite{Holm1985}. Likewise, if $||\cdot||:\mathfrak{X}(\mathcal{D})\times V^*$ is any norm such that the linear mapping $\mathfrak{C}:\mathfrak{X}(\mathcal{D})\times V\rightarrow \mathfrak{X}(\mathcal{D})\times V$ is bounded, the norm of the evolution of a dynamically accessible $(\delta\mathbf{v}_t,\delta a_t)$, $||(\delta\mathbf{v}_t,\delta a_t)||\leq||\mathfrak{C}||\cdot ||(\bm{\xi}_t,u_t)||$, will be bounded for all time when $\delta^2 h$ is positive definite, thus implying stability of (linear) dynamically accessible perturbations. Note that $\delta^2 h(\bm{\xi},u)$ \emph{does not} directly define a norm on the space of dynamically accessible variations when it is positive definite because Eq.\,(\ref{D2HEP}) is not a quadratic functional of $(\delta\mathbf{v},\delta a)$.

\section{An elementary derivation of the dynamically accessible stability criterion}\label{energy}
Now we will apply the general theory developed in the previous two sections to relative equilibria of the ideal MHD equations. First, we will cast the nonlinear MHD equations in the form of an Euler-Poincar\'e system using the Lagrangian
\begin{align}
&L[\mathbf{x},\mathbf{v},\rho_o\,\mathrm{d}^3x,s_o,\mathbf{B}_o\cdot\mathrm{d}\mathbf{S}]=\\
&\int_\mathcal{D}\left(\frac{1}{2}\mathbf{v}\cdot\mathbf{v}-\epsilon\left(\mathbf{x}^{-1*}s_o,\mathcal{J}(\mathbf{x}^{-1})\mathbf{x}^{-1*}\rho_o\right)\right)\mathbf{x}^{-1*}(\rho_o\mathrm{d}^3x)\nonumber\\
&-\frac{1}{2}\int_\mathcal{D}\mathcal{J}(\mathbf{x}^{-1})^2(\mathbf{x}^{-1*}\mathbf{B}_o)\cdot (\mathbf{x}^{-1*}\mathbf{B}_o)\,\mathrm{d}^3x\nonumber
\end{align} 
where $\mathcal{J}(\mathbf{x})\,\mathrm{d}^3x\equiv\mathbf{x}^*\,\mathrm{d}^3x$ defines the Jacobian of the mapping $\mathbf{x}$, $a\equiv(\rho_o\,\mathrm{d}^3x,s_o,\mathbf{B}_o\cdot\mathrm{d}\mathbf{S})$ are the advected parameters, and the specific internal energy $\epsilon$ is defined by the first law of thermodynamics $\mathrm{d}\epsilon=T(s,\rho)\mathrm{d}s-p(s,\rho)\mathrm{d}(1/\rho)$, with $T$ and $p$ being the temperature and the pressure respectively.
The advected quantities correspond to the local conservation (advection) of mass, entropy, and magnetic flux respectively. 

It is straightforward to verify that this Lagrangian satisfies the symmetry property characteristic of Euler-Poincar\'e fluids, $L(\mathbf{x}\circ\mathbf{y},\mathbf{v},\mathbf{y}^*a)=L(\mathbf{x},\mathbf{v},a)$. Therefore, the reduced Lagrangian can be calculated, giving
\begin{align}
l(\mathbf{v},a)=\int_\mathcal{D}\left(\frac{1}{2}\mathbf{v}\cdot\mathbf{v}-\epsilon(s,\rho)\right)\rho\,\mathrm{d}^3x-\frac{1}{2}\int_\mathcal{D}\mathbf{B}\cdot\mathbf{B}\,\mathrm{d}^3x,
\end{align} 
with $a=(\rho\,\mathrm{d}^3x,s,\mathbf{B}\cdot \mathrm{d}\mathbf{S})$. Using the constrained variations
\begin{eqnarray}
\delta\mathbf{v}&=&\dot{\bm{\xi}}+\mathbf{v}\cdot\nabla\bm{\xi}-\bm{\xi}\cdot\nabla\mathbf{v},\label{VELMHD}\\
\delta\rho&=&-\nabla\cdot(\rho\bm{\xi}),\label{Vmass1}\\
\delta s&=&-\bm{\xi}\cdot\nabla s,\label{Ventropy1}\\
\delta\mathbf{B}&=&\nabla\times(\bm{\xi}\times\mathbf{B}),\label{Vflux1}
\end{eqnarray}
which can be derived from Eq.\,(\ref{constrained_variations}), the reduced variational principle $\delta\int l\,\mathrm{d}t=0$ then leads to the momentum equation,
\begin{equation}
\frac{\mathrm{d}}{\mathrm{d}t}(\rho\mathbf{v})=-\nabla\cdot(\rho\mathbf{v}\mathbf{v})+(\nabla\times\mathbf{B})\times\mathbf{B}-\nabla p,\label{Tmomentum}
\end{equation}
which corresponds to Eq.\,(\ref{TEP}) with $\delta L/\delta \mathbf{v}=\rho\mathbf{v}\otimes\mathrm{d}^3x$. This equation, together with the analogue of Eq.\,(\ref{Tadvected}),
\begin{eqnarray}
\frac{\mathrm{d}}{\mathrm{d}t}\rho&=&-\nabla\cdot(\rho\mathbf{v}),\label{Tmass}\\
\frac{\mathrm{d}}{\mathrm{d}t} s&=&-\mathbf{v}\cdot\nabla s,\label{Tentropy}\\
\frac{\mathrm{d}}{\mathrm{d}t}\mathbf{B}&=&\nabla\times(\mathbf{v}\times\mathbf{B}),\label{Tflux}
\end{eqnarray}
then gives the complete ideal MHD equations as an Euler-Poincar\'e system. Such a formulation was first presented by Newcomb \cite{Newcomb}.

Using this formulation, we will now present Frieman and Rotenberg's stability criterion \cite{Frieman} for a relative equilibrium $(\mathbf{v}_o,a_o)$. First, we linearize the ideal MHD equations (\ref{Tmomentum})-(\ref{Tflux}) and obtain the linearized MHD equations for perturbations $(\delta\mathbf{v}, \delta a)$. Then, by expressing $(\delta\mathbf{v}, \delta a)$ in terms of $(\bm{\xi}, \dot{\bm{\xi}})$ using Eqs.\,(\ref{VELMHD})-(\ref{Vflux1}), we have the linearized equation of motion \cite{Frieman}
\begin{align}
\rho_o\ddot{\bm{\xi}}+2\rho_o \mathbf{v}_o\cdot\nabla\dot{\bm{\xi}}=\mathbf{F}(\bm{\xi}),\label{Txi2}
\end{align}
where
\begin{align}
\mathbf{F}(\bm{\xi})=\nabla\cdot(\rho_o\bm{\xi}\mathbf{v}_o\cdot\nabla\mathbf{v}_o-\rho_o\mathbf{v}_o\mathbf{v}_o\cdot\nabla\bm{\xi})+\mathbf{F}_0(\bm{\xi}),
\end{align}
and $\mathbf{F}_0$ is the linearized force operator for equilibria without flow \cite{Freidberg}. Using Eq.\,(\ref{Txi2}) and the self-adjointness of $\mathbf{F}$, it is simple to verify that the quantity 
\begin{align}
H_2(\bm{\xi}, \dot{\bm{\xi}})=\int_\mathcal{D}\dot{\bm{\xi}}\cdot\dot{\bm{\xi}}\rho_o\,\mathrm{d}^3x-\int_\mathcal{D}\bm{\xi}\cdot\mathbf{F}(\bm{\xi})\,\mathrm{d}^3x
\end{align}
is conserved \cite{Frieman,Hameiri}. A simple argument \cite{Freidberg} then shows that this implies the relative equilibrium is stable when $\delta^2W(\bm{\xi})=-\int\bm{\xi}\cdot\mathbf{F}(\bm{\xi})\,\mathrm{d}^3x$ is positive definite. This is exactly Frieman and Rotenberg's sufficient condition for linear stability. 

Next, we will derive Hameiri's sufficient condition for linear stability of a relative equilibrium against dynamically accessible perturbations. Our method is essentially that discussed earlier in the general context of Euler-Poincar\'e fluids, but all of the manipulations we will perform can be verified easily without recourse to the abstract theory. Consider a linear perturbation $(\delta\mathbf{v}, \delta a)$ to equilibrium $(\mathbf{v}_o,a_o)$ of the dynamically accessible form \cite{Hameiri,Morrison}
\begin{align} 
\label{Vvelocity}\delta\mathbf{v}&=\bm{\xi}\times(\nabla\times\mathbf{v}_o)+{(\nabla\times\bm{\zeta})\times\mathbf{B}_o}/{\rho_o}+\alpha\nabla s_o-\nabla\beta\\
\label{Vmass}\delta\rho&=-\nabla\cdot(\rho_o\bm{\xi})\\
\label{Ventropy}\delta s&=-\bm{\xi}\cdot\nabla s_o\\
\label{end}\delta\mathbf{B}&=\nabla\times(\bm{\xi}\times\mathbf{B}_o)
\end{align}
and regard it as an initial condition to the MHD equations linearized about the equilibrium.

First we will show that, as this initial perturbation evolves under the linearized MHD equations, it will remain in this form. To see this, suppose that $(\bm{\xi},\bm{\zeta},\alpha,\beta)$ satisfied the evolution equations
\begin{align}
\label{l1}\dot{\bm{\xi}}+\mathbf{v}_o\cdot\nabla\bm{\xi}-\bm{\xi}\cdot\nabla\mathbf{v}_o&=\delta\mathbf{v}\\
\dot{\beta}+\mathbf{v}_o\cdot\nabla\beta&={\delta p}/{\rho_o}+\bm{\xi}\cdot{(\nabla\times\mathbf{B}_o)\times\mathbf{B}_o}/{\rho_o}\label{Dmass}\\
\dot{\alpha}+\mathbf{v}_o\cdot\nabla\alpha&=\delta T+\bm{\xi} \cdot\nabla T_o\label{Dentropy}\\
\dot{\bm{\zeta}}+\mathbf{v}_o\cdot\nabla\bm{\zeta}+\nabla{\mathbf{v}_o}\cdot\bm{\zeta}&=\delta\mathbf{B}+\bm{\xi}\cdot\nabla\mathbf{B}_o+\nabla\bm{\xi}\cdot\mathbf{B}_o,\label{Dflux}
\end{align}
where every instance of $\delta\mathbf{v}$ or $\delta a$ is to be expressed in terms of $(\bm{\xi},\bm{\zeta},\alpha,\beta)$ using Eqs.\,(\ref{Vvelocity})-(\ref{end}). Then it is a straightforward exercise in algebra to verify that the resulting $(\delta\mathbf{v}_t,\delta a_t)$ satisfies the linearized MHD equations. But this $(\delta\mathbf{v}_t,\delta a_t)$ is obviously in the form specified by Eqs.\,(\ref{Vvelocity})-(\ref{end}), which verifies the claim because of uniqueness of solutions to the linearized equations.

Next, we will derive a condition for stability of the solutions to the evolution equation specified in Eqs.\,(\ref{l1})-(\ref{Dflux}). As a first step, notice that the variable $\bm{\xi}$ satisfies Eq.\,(\ref{Txi2}). Thus, $H_2(\mathbf{\xi},\dot{\mathbf{\xi}})$ is constant in time. But this means that the quadratic functional
\begin{align}\label{D2HDA}
&\delta^2h(\bm{\xi},\bm{\zeta},\alpha,\beta)=\\
&\int_\mathcal{D}(\delta\mathbf{v}-\mathbf{v}_o\cdot\nabla\bm{\xi}+\bm{\xi}\cdot\nabla\mathbf{v}_o)^2\rho_o\,\mathrm{d}^3x-\int_\mathcal{D}\bm{\xi}\cdot\mathbf{F}(\bm{\xi})\,\mathrm{d}^3x,\nonumber
\end{align}
where $\delta\mathbf{v}$ is expressed in terms of $(\bm{\xi},\bm{\zeta},\alpha,\beta)$ using Eq.\,(\ref{Vvelocity}), is conserved along solutions to the evolution equations (\ref{l1})-(\ref{Dflux}). Therefore, whenever the quadratic functional $\delta^2h$ is positive definite, it defines a norm that bounds solutions to the evolution equations for $(\bm{\xi},\bm{\zeta},\alpha,\beta)$ for all time, which implies stability.

Finally, we can give a sufficient condition for stability of the relative equilibrium $(\mathbf{v}_o,a_o)$ to linear perturbations of the form of Eqs.\,(\ref{Vvelocity})-(\ref{end}). When the functional $\delta^2h$ is positive definite, we have just found that solutions to the evolution equations for $(\bm{\xi},\bm{\zeta},\alpha,\beta)$ are bounded for all time by the norm defined by $\delta^2h$. But we also know that every solution to the linearized MHD equations with this type of initial condition can be expressed in terms of a solution to the $(\bm{\xi},\bm{\zeta},\alpha,\beta)$ evolution equations using Eqs.\,(\ref{Vvelocity})-(\ref{end}). Therefore, given any norm on the space of $(\delta\mathbf{v},\delta a)$'s such that the linear mapping specified by Eqs.\,(\ref{Vvelocity})-(\ref{end}) is bounded, $(\delta\mathbf{v}_t,a_t)$ will be bounded for all time in that norm. That is to say, whenever $\delta^2 h$ is positive definite, linear perturbations to the relative equilibrium $(\mathbf{v}_o,a_o)$ of the form specified by Eqs.\,(\ref{Vvelocity})-(\ref{end}) will be stable. This is precisely the condition derived by Hameiri\cite{Hameiri}. Clearly, it only directly applies to a restricted class of the initial perturbations appearing in Frieman and Rotenberg's stability criterion.

Note that only simple algebra, coupled with an apparently lucky guess for the appropriate evolution equations for $(\bm{\xi},\bm{\zeta},\alpha,\beta)$, was necessary to derive this condition. In order to understand how the evolution equations for $(\bm{\xi},\bm{\zeta},\alpha,\beta)$ were deduced, one should refer to the general theory for Euler-Poincar\'e fluids developed in the previous section; the evolution equations follow from Eqs.\,(\ref{VEL}) and (\ref{DEP}). Also, it is not a coincidence the quadratic functional $\delta^2h$ is conserved, as it can be obtained directly \cite{Hameiri} from Eq.\,(\ref{D2HEP2}), with its conservation guaranteed by Eq.\,(\ref{conserve}). The self-adjointness of $\mathbf{F}(\bm{\xi})$ can, in turn, be proved as a result of the conservation of $\delta^2h$.

Moreover, in light of the general theory, the special form of perturbations considered in Hameiri's stability criterion are the most general linear perturbations that preserve the Noether invariants implied by the particle relabeling symmetry. Indeed, Eqs.\,(\ref{Vvelocity})-(\ref{end}) follow from the expression for the mapping $\mathfrak{C}$ defined in Eq.\,(\ref{c}). On the other  hand, from the Poisson perspective used by Hameiri \cite{Hameiri} and Morrison \cite{Morrison}, this type of perturbation was found to be the most general form of linear perturbation that preserves the Casimir invariants. Thus, we conclude that in order to study dynamically accessible perturbations to systems that admit both a Poisson and an Euler-Poincar\'e formulation, one can either use the Poisson formulation to find perturbations that conserve the Casimirs, or the Euler-Poincar\'e formulation to find perturbations that preserve the Noether invariants implied by particle relabeling symmetry.

\section{Discussion}\label{discussion}
Eqs.\,(\ref{Vvelocity})-(\ref{end}) were previously obtained using the non-canonical Poisson bracket for the ideal MHD and referred to as dynamically accessible variations by Morrison\cite{Morrison}. They were constructed to be the most general linear perturbations that preserve all Casimirs of the system. From the point of view adopted in this paper, this form of perturbation follows from Eq.\,(\ref{c}), and therefore gives the most general form of linear perturbation that preserves the Noether invariants implied by the particle relabeling symmetry \cite{Cotter} and the advected parameters. We expect that this correspondence can be proven to hold more generally in systems that admit both a Poisson and Euler-Poincar\'e description using the fact the Poisson formulation can be derived from the Euler-Poincar\'e formulation using the Legendre transform\cite{book}.

Using Eqs.\,(\ref{Vvelocity})-(\ref{end}) as constraints, Hameiri \cite{Hameiri} obtained $\delta^2H$ as in Eq.\,(\ref{D2H}) with $\delta\mathbf{v}$ expressed in terms of $(\bm{\xi}, {\bm{\zeta}}, {\alpha}, \beta)$ using Eq.\,(\ref{VDA}). Hameiri then claimed that positivity of $\delta^2H$ as a functional of $(\bm{\xi}, {\bm{\zeta}}, {\alpha}, \beta)$ was sufficient for linear stability of the equilibrium to dynamically accessible variations. The justification of this claim was not clearly explained. In fact, the justification can be deduced from relevant discussions in Refs.\,\onlinecite{Hameiri,Morrison}. Alternatively, the argument presented in this paper provides another proof. More importantly, our derivation clearly shows that this criterion only applies to perturbations that preserve the Noether invariants.

As mentioned in the introduction, Hameiri claimed that his stability criterion is stronger than what Frieman and Rotenberg obtained \cite{Frieman} via direct analysis of the linearized equations of motion (\ref{Txi}). But in this paper, we have shown that such a claim is incorrect because the two stability criteria apply to different types of perturbations. In the derivation of Frieman and Rotenberg's criterion, only Eqs.\,(\ref{Vmass})-(\ref{l1}) were used, so the perturbations preserve the advected parameters but not necessarily the Noether invariants, which is in line with the traditional energy principle for absolute equilibria \cite{Freidberg,Bernstein}. On the other hand, Hameiri's criterion further requires the initial $\delta\mathbf{v}$ to be of the form of Eq.\,(\ref{Vvelocity}) to preserve the Noether invariants. However, if one only considers the dynamically accessible perturbations, then Frieman and Rotenberg's condition is indeed weaker than Hameiri's condition, i.e. the Hameiri condition could indicate stability against dynamically accessible perturbations while the Frieman and Rotenberg condition is inconclusive. It should also be noted that for initial perturbations that do not conserve the advected parameters, the applicability of both criteria, and even the traditional energy principle, becomes questionable.

Finally, we will conclude with a summary of our results. By presenting and comparing two existing stability criteria for ideal MHD equilibria with flow within the same framework of the Euler-Poincar\'e theory, we have shown that they actually apply to different types of perturbations. We also have formulated a Lagrangian analogue to the method of dynamically accessible perturbations for stability analysis that applies to general Euler-Poincar\'e fluids.

\begin{acknowledgements}
The authors would like to thank C. Liu and J. Squire for helpful discussions. This work was supported by the U.S. Department of Energy under Contract No.\,DE-AC02-09CH11466.
\end{acknowledgements}

\appendix
\section{Some Notation}
\emph{One-form densities}\cite{Holm,book}: The dual space associated to $\mathfrak{X}(\mathcal{D})$, $\mathfrak{X}(\mathcal{D})^*$, is naturally identified with the set of one-form densities, $\mathfrak{X}(\mathcal{D})^*=\Gamma(\Omega^1(\mathcal{D})\otimes\Omega^3(\mathcal{D}))$. Thus, a typical element of $\mathfrak{X}(\mathcal{D})^*$ assigns to each point $X\in \mathcal{D}$ a tensor of the form $\theta(X)\otimes\sigma_o$, where $\theta$ is some one-form and $\sigma_o=\mathrm{d}^3x$ is the (constant) standard volume form on $\mathcal{D}$. Given a vector field $\mathbf{w}$ on $\mathcal{D}$, the Lie derivative of such a one-form density is given by $\mathfrak{L}_{\mathbf{w}}(\theta\otimes\sigma_o)=(\mathfrak{L}_{\mathbf{w}}\theta)\otimes\sigma_o+\theta\otimes(\mathfrak{L}_{\mathbf{w}}\sigma_o)=(\mathfrak{L}_{\mathbf{w}}\theta+\theta\nabla\cdot\mathbf{w})\otimes\sigma_o$. If we denote the pairing between a vector field and a one-form density as $\left<\theta\otimes\sigma_o,\mathbf{w}\right>\equiv\int\theta(\mathbf{w})\sigma_o$, we then see that $\left<\theta\otimes\sigma_o,\mathfrak{L}_{\mathbf{v}}\mathbf{w}\right>=-\left<\mathfrak{L}_{\mathbf{v}}(\theta\otimes\sigma_o),\mathbf{w}\right>$. Note that, in order to derive this identity, one must utilize the fact that vector fields are tangent to the boundary of $\mathcal{D}$.

\emph{Functional derivatives}: Given a functional $l:\mathfrak{X}(\mathcal{D})\times V^*\rightarrow\mathbb{R}$, its partial functional derivatives are defined by the equation
\begin{align}
\frac{\mathrm{d}}{\mathrm{d}\epsilon}\bigg|_0&l(\mathbf{v}+\epsilon\delta\mathbf{v},a+\epsilon\delta a)=\\
&\left<\frac{\delta l}{\delta\mathbf{v}}(\mathbf{v},a),\delta\mathbf{v}\right>+\left<\delta a,\frac{\delta l}{\delta a}(\mathbf{v},a)\right>.\nonumber
\end{align} 
Here, $\left<\cdot,\cdot\right>$ denotes the pairing between a vector space and its dual and $(\delta\mathbf{v},\delta a)\in\mathfrak{X}(\mathcal{D})\times V^*$. We therefore see that for each $(\mathbf{v},a)$, $\frac{\delta l}{\delta\mathbf{v}}(\mathbf{v},a)\in\mathfrak{X}(\mathcal{D})^*$ is a one-form density and $\frac{\delta l}{\delta a}(\mathbf{v},a)\in V^{**}=V$ is a $k$-form.

We will also have occasion to use higher functional derivatives in this article. Specifically, in Eq.\,(\ref{D2HEP2}), we employ the notations $\frac{\delta^2l}{\delta a\delta\mathbf{v}},\frac{\delta^2l}{\delta\mathbf{v}\delta a},\frac{\delta^2}{\delta a^2},$ and $\frac{\delta^2l}{\delta\mathbf{v}^2}$. The relevant definitions are
\begin{align}
\frac{\delta^2l}{\delta a\delta\mathbf{v}}(\mathbf{v},a)[\delta a]&=\frac{\mathrm{d}}{\mathrm{d}\epsilon}\bigg|_0\frac{\delta l}{\delta\mathbf{v}}(\mathbf{v},a+\epsilon\delta a)\\
\frac{\delta^2l}{\delta\mathbf{v}\delta a}(\mathbf{v},a)[\delta \mathbf{v}]&=\frac{\mathrm{d}}{\mathrm{d}\epsilon}\bigg|_0\frac{\delta l}{\delta a}(\mathbf{v}+\epsilon\delta\mathbf{v},a)\\
\frac{\delta^2l}{\delta a^2}(\mathbf{v},a)[\delta a]&=\frac{\mathrm{d}}{\mathrm{d}\epsilon}\bigg|_0\frac{\delta l}{\delta a}(\mathbf{v},a+\epsilon\delta a)\\
\frac{\delta^2l}{\delta \mathbf{v}^2}(\mathbf{v},a)[\delta \mathbf{v}]&=\frac{\mathrm{d}}{\mathrm{d}\epsilon}\bigg|_0\frac{\delta l}{\delta \mathbf{v}}(\mathbf{v}+\epsilon\delta\mathbf{v},a).
\end{align}
An unfortunate consequence of this notation is that $\frac{\delta^2l}{\delta a\delta\mathbf{v}}(\mathbf{v},a)\neq\frac{\delta^2l}{\delta\mathbf{v}\delta a}(\mathbf{v},a)$. That this is true is obvious from the above definitions; $\frac{\delta^2l}{\delta a\delta\mathbf{v}}(\mathbf{v},a)$ is linear map from $V^*$ into $\mathfrak{X}(\mathcal{D})^*$ while $\frac{\delta^2l}{\delta\mathbf{v}\delta a}(\mathbf{v},a)$ is a linear map from $\mathfrak{X}(\mathcal{D})$ into $V$.

\emph{Diamond product}\cite{Holm,book}: Given $(u,a)\in V\times V^*$, consider the linear functional on $\mathfrak{X}(\mathcal{D})$, $\mathbf{w}\mapsto\left<\mathfrak{L}_{\mathbf{w}}a,u\right>$. By definition, this linear functional specifies an element, $\tilde{\theta}(u,a)$, of the dual space $\mathfrak{X}(\mathcal{D})^*$ that satisfies
\begin{align}
\left<\mathfrak{L}_{\mathbf{w}}a,u\right>=\left<\tilde{\theta}(u,a),\mathbf{w}\right>.
\end{align} 
Note that on the left-hand side of this expression, the pairing is between $V$ and $V^*$, while on the right-hand side it is between $\mathfrak{X}(\mathcal{D})$ and $\mathfrak{X}(\mathcal{D})^*$. Because $\tilde{\theta}(u,a)$ must be a bilinear function of its arguments, it defines a $\mathfrak{X}(\mathcal{D})^*$-valued product $(u,a)\mapsto u\diamond a$ that satisfies
\begin{align}
\left<u\diamond a,\mathbf{w}\right>=-\left<\mathfrak{L}_{\mathbf{w}} a,u\right>.
\end{align} 

\end{document}